\documentclass[12pt,a4paper]{article}
\usepackage{a4wide}
\usepackage{amsmath,amssymb,amsfonts,amsxtra,simpler-wick} 
\usepackage{cite}

\allowdisplaybreaks

\begin{document}
\thispagestyle{empty}

\begin{center}{\Large \bf Lyapunov exponents in $\mathbf{{\cal N}=2}$
    supersymmetric 
    Jackiw-Teitelboim 
  gravity}\end{center}  

\vspace*{1cm}

\centerline{Ruben Campos Delgado and Stefan F{\"o}rste }

\vspace{1cm}

\begin{center}{\it
Bethe Center for Theoretical Physics\\
{\footnotesize and}\\
Physikalisches Institut der Universit\"at Bonn,\\
Nussallee 12, 53115 Bonn, Germany}
\end{center}

\vspace*{1cm}

{\let\thefootnote\relax\footnotetext{ruben.camposdelgado@gmail.com, \,
  forste@th.physik.uni-bonn.de}} 

\centerline{\bf Abstract}
\vskip .3cm

We study $\mathcal{N}=2$ supersymmetric Jackiw-Teitelboim (JT) gravity
at finite temperature coupled to matter. 
The matter fields are related to superconformal primaries by AdS/CFT
duality. 
Due to broken super reparametrisation invariance in the SCFT dual,
there are corrections to superconformal correlators. These are
generated by the exchange of super-Schwarzian modes which is
dual to the exchange of 2D supergravity modes.
We compute corrections to four-point
functions for superconformal primaries and
analyse the behaviour of out-of-time-ordered correlators.
In particular, four-point functions of two pairs of primaries with
mutually vanishing two-point functions are considered.
By decomposing the corresponding supermultiplet into its components, we 
find different Lyapunov exponents. The value of the Lyapunov exponents
depends on 
whether the correction is due to graviton, gravitini or graviphoton
exchange. If mutual two-point functions do not vanish all
components grow with maximal Lyapunov exponent.


\vskip .3cm

\newpage
\section{Introduction}
Lower dimensional instances of the AdS$_D$/CFT$_{D-1}$ correspondence
\cite{Maldacena:1997re} have attracted a wide interest in recent
years. The case $D = 2$ is important from the point 
of view of black hole physics because AdS$_2$ is the near-horizon
geometry of extremal (e.g.\ supersymmetric) black holes with non-zero
entropy. A known 
problem is that pure gravity in AdS$_2$ is inconsistent with the
existence of finite energy excitations above the AdS$_2$ vacuum
\cite{Fiola:1994ir, Maldacena:1998uz, Almheiri:2014cka}. The problem
can be overcome by considering a deformed version of the AdS$_2$
space.  
In Euclidean signature, AdS$_2$ is the hyperbolic disk. The idea is to
slightly deform the boundary of the disk by means of a UV brane, i.e.\
a curve that reduces to the old boundary as soon as a small parameter
is sent to zero, without changing the AdS$_2$ geometry. In this nearly
AdS$_2$ space (nAdS$_2$) it is possible to define a consistent
2-dimensional theory of gravity, the Jackiw-Teitelboim (JT) theory
\cite{Jackiw:1984je, Teitelboim:1983ux}, coupled to a dilaton
field. The dilaton field enters as a Lagrange multiplier enforcing
negative constant curvature. After imposing the constraint, JT
gravity reduces to an 
effective action involving the Schwarzian derivative of a
reparametrisation of the coordinate describing the nAdS$_2$ boundary
\cite{Turiaci:2016cvo, Jensen:2016pah, Maldacena:2016upp,
  Cvetic:2016eiv}: 
\begin{equation}\label{eq:bosonic_schw}
    S_{\text{sch}}=-C\int du\,\text{Sch}\left[t(u);u\right],
\end{equation}
where $C$ is inversely proportinal to Newton's constant $G_N$ and
\begin{equation}
    \text{Sch}\left[t(u);u\right]=\frac{2\dot{t}t^{(3)}-3\ddot{t\,}^2}{2\dot{t\,}^2}.
\end{equation}
The JT theory has surprisingly some features in common with a
completely different model, originally studied in condensed matter
physics, the SYK model \cite{Sachdev:1992fk, Kitaev:2015m27}. It
describes $N$ fermions interacting via random couplings.  The model is
solvable in the large $N$ limit, is chaotic and is conformally
invariant in the low energy regime \cite{Kitaev:2015m27,
  Parcollet:1999nf, Sachdev:2010um, Sachdev:2010uj}. Moving away from
the IR region breaks the conformal symmetry, resulting in an effective
action for time reparametrisations given again by the Schwarzian.   
This corroborates JT gravity as the holographic dual of the SYK
low-energy region. To be precise, this is an example of a nearly
AdS$_2$/nearly CFT$_1$ correspondence. In the same spirit of the usual
``holographic dictionary" of the canonical AdS/CFT correspondence
\cite{Witten:1998qj}, matter fields in nAdS$_2$ correspond  to
conformal primaries in the conformal limit of the SYK model. More
precisely, correlation functions of conformal primaries are obtained
by taking variational derivatives of the gravity partition function
with respect to boundary conditions of matter fields. The breaking of
conformal symmetry is characterised by a Schwarzian action for
reparametrisations of the boundary. These are responsible for
corrections to correlation functions signaling the occurence of
quantum chaos.
In a chaotic system, small differences in the initial conditions lead
to large and unpredictable differences in the dynamics. In a quantum
theory, a diagnosis of chaos is offered by the behaviour of a special
class of correlators, the out-of-time-ordered correlators (OTOC). If
the system is chaotic they show an exponential behaviour with rate
given by a real number $\lambda$ called Lyapunov exponent.  
For a causal and unitary QFT it was argued that the Lyapunov exponent
is bounded by a maximal value \cite{Maldacena:2015waa}. Calculations
carried in JT gravity \cite{Maldacena:2016upp} and in the SYK model
\cite{Kitaev:2015m27, Maldacena:2016hyu} gave the maximal value of
Lyapunov exponent for both theories, thereby supporting the duality
between them.  

Both the JT theory and the SYK model can be generalised by adding
supersymmetry. In \cite{Fu:2016vas} it was shown that the
$\mathcal{N}=1$ and $\mathcal{N}=2$ supersymmetric versions of the SYK
model lead to super-Schwarzian actions. Similarly to the purely
bosonic theory, the super-Schwarzian also arises as an effective
action for a UV regulator brane in nAdS$_2$ supergravity, for both the
$\mathcal{N}=1$ \cite{Forste:2017kwy} and $\mathcal{N}=2$
\cite{Forste:2017apw} supersymmetric JT theories. This suggests that
the JT/SYK duality is still preserved after adding supersymmetry. 

Lyapunov exponents were considered for ${\cal N}=1$ JT super
gravity in \cite{Narayan:2017hvh,Mertens:2020pfe}. For ${\cal N} =2$
they were computed in \cite{Bulycheva:2018qcp} within the SYK model. Here, we
carry out the holographic computation for ${\cal N}=2$ and obtain
results similar to \cite{Narayan:2017hvh}.
Whenever the leading correction to the four-point functions contains 
graviton exchange the Lyapunov exponent saturates the bound
$2\pi/\beta$. There are also components of the four-point functions for
which the leading correction does not contain graviton exchange. If
gravitino exchange contributes, their Lyapunov exponent will be
$\pi/\beta$ whereas for combinations with graviphoton contributions
only there will be just linear growth with time. 

We briefly outline the procedure that we are going to follow. Firstly,
we put the theory at finite temperature and compute the corresponding
super-Schwarzian action. Secondly, we expand the reparametrisations of
the super-boundary in terms of small fluctuations. Subsequently, we
put matter fields into the nAdS$_2$ space and compute the four-point
functions of the dual operators, contracting the appearances of the
fluctuations using their propagators.  Using its definition, we are
eventually able to extract the Lyapunov exponents from the four-point
functions. The same procedure was carried out for the purely bosonic
JT theory in \cite{Maldacena:2016upp} and reviewed in
\cite{Sarosi:2017ykf}. 
The paper is organised as follows. In Section
\ref{sec:N=2_formalism} we review the $\mathcal{N}=2$ superspace
formalism and compute the super-Schwarzian at finite temperature. In
Section \ref{sec:propagators} we derive the propagators of the
fluctuations. In Section \ref{sec:N=2matter} we couple the theory to
matter. In Section \ref{sec:N=2_4pointfunc} 
we compute the thermal four-point functions for operators dual to the
matter fields and determine
the Lyapunov exponents. 
\section{The $\mathcal{N}=2$ superspace formalism  }\label{sec:N=2_formalism} 
In the purely bosonic JT theory, the field $t(u)$ is
the only dynamical variable describing reparametrisations of the
boundary in the nAdS$_2$ 
space. In a supersymmetric theory, we have to add one or more
fermionic fields. The superspace formalism offers a simple way to keep
track of all the relevant degrees of freedom and to make supersymmetry
manifest. Within this framework, space is described in terms of the
usual bosonic coordinates and by additional Grassmann variables. In a
generic $\mathcal{N}=m$ theory, super-reparametrisations of the
boundary  are described by an $(m+1)$-tuple  
$\left(t,\xi_1,\dots,\xi_m\right)$ where $t$ respectively $\xi_i$ is a
commuting respectively anti-commuting superfield satisfying additional
constraints which can be viewed as  one-dimensional versions of
superconformal Killing equations.    
\subsection{$\mathcal{N}=2$ superspace and super-Schwarzian}  
We briefly review definitions of ${\cal N}=2$ super-diffeomorphisms
following conventions in \cite{Fu:2016vas}.
The $\mathcal{N}=2$ one-dimensional superspace, or super-line, is
parametrised by a bosonic variable $u$ and Grassmann variables
$\theta$ and $\overline{\theta}$. The super-translation group consists of
the transformations  
\begin{equation}\label{eq:N=2translations} 
    u\mapsto\tau=u+\epsilon+\theta\overline{\eta}+\overline{\theta}\eta,
    \hspace{4mm}
    \theta\mapsto\xi=\theta+\eta,
    \hspace{4mm}\overline{\theta}\mapsto\overline{\xi}=\overline{\theta}+\overline{\eta}.  
\end{equation}
where $\epsilon(u)$ is a bosonic function and $\eta(u)$,
$\overline{\eta}(u)$ are fermionic functions. 
Super-derivative operators are defined as
\begin{equation}
  D_{\theta}=\frac{\partial}{\partial\theta}+\overline{\theta}
  \frac{\partial}{\partial u}, \hspace{6mm} \overline{D}_{\overline{\theta}}=
  \frac{\partial}{\partial\overline{\theta}}+\theta\frac{\partial}{\partial u}.
\end{equation}
Their anticommutator is
\begin{equation}\label{eq:anticommutator_derivatives}
    \{D_{\theta}, \overline{D}_{\overline{\theta}}\}=2\partial_u.
\end{equation}
Fermions and scalars can be packaged into a chiral superfield,
i.e.\ a field $\Psi\left(u,\theta, \overline{\theta}\right)$ constrained to satisfy
\begin{equation}
    \overline{D}_{\overline{\theta}}\Psi\left(u,\theta,\overline{\theta}\right)=0,
\end{equation}
which is solved by 
\begin{equation}
  \Psi\left(u,\theta,\overline{\theta}\right)=
  \Gamma\left(u+\theta\overline{\theta}\right)+\theta
    b(u) 
\end{equation}
for a fermion $\Gamma$ and boson $b$. 
The conjugate $\overline{\Psi}$ and $D_{\theta}\Psi$ are instead
anti-chiral field, namely they are annihilated by $D_{\theta}$.  

Transormations in (\ref{eq:N=2translations}) are
super-reparametrisations if they satisfy the constraints
\begin{equation}\label{eq:N=2constraints}
\begin{split}
&D_{\theta}\overline{\xi}=0, \hspace{8 mm} D_{\theta}\tau=\overline{\xi}D_{\theta}\xi, \\
&\overline{D}_{\overline{\theta}}\xi=0, \hspace{8 mm} \overline{D}_{\overline{\theta}}\tau=\xi \overline{D}_{\overline{\theta}}\overline{\xi}.
\end{split}
\end{equation}
Super-reparametrisations are described by an effective action which
can be obtained from a supersymmetric SYK model \cite{Fu:2016vas}  as
well as from supersymmetric JT gravity 
\cite{Forste:2017apw}
\begin{equation}\label{eq:N=2superschwarzian_action}
    S_{\text{eff}}=-\frac{1}{2\pi G_N}\int_{\partial M}du d\theta d\overline{\theta}\,\phi_b \text{Sch}\left[\tau,\xi,\overline{\xi};u,\theta,\overline{\theta}\right],
\end{equation}
where $\phi_b$ is the boundary value for the leading component of the dilaton
$\Phi$. For convenience we rename 
\begin{equation}
   \kappa=-\frac{\phi_b}{2\pi G_N}   
 \end{equation}
 and consider constant $\phi_b$ which can be achieved by an
 appropriate choice of the boundary. 
The functional
$\text{Sch}\left[\tau,\xi,\overline{\xi};u,\theta,\overline{\theta}\right]$ is
the $\mathcal{N}=2$ super-Schwarzian derivative \cite{Fu:2016vas} 
\begin{equation}\label{eq:N=2schw_derivative}
    \text{Sch}\left[\tau,\xi,\overline{\xi};u,\theta,\overline{\theta}\right]=\frac{\partial_u\overline{D}_{\overline{\theta}}\overline{\xi}}{\overline{D}_{\overline{\theta}}\overline{\xi}}-\frac{\partial_u D_{\theta}\xi}{D_{\theta}\xi}-2\frac{\partial_u\xi\partial_u\overline{\xi}}{\left(D_{\theta}\xi\right)\left(\overline{D}_{\overline{\theta}}\overline{\xi}\right)},
\end{equation}
where $\tau$, $\xi$, $\overline{\xi}$ are super-reparametrisations of the boundary of nAdS$_2$ satisfying the constraints \eqref{eq:N=2constraints}. 
\subsection{The $\mathcal{N}=2$ super-Schwarzian at finite temperature }\label{sec:N=2fin_temp}
We now put the theory at finite temperature $T=1/\beta$. This is
achieved by reparametrising the boundary time to a tangent,
$\tau\left(u\right)
\mapsto\tan{\frac{2\pi}{\beta}\frac{\tau\left(u\right)}{2}}$ 
\cite{Maldacena:2016upp}.  For
simplicity we fix $\beta= 2\pi$. Assuming that $\left( \tau, \xi, \overline{\xi}\right)$
  are super-reparametrisations it is easy to check that 
\begin{equation}\label{eq:thermal_rep}
\tau'=\tan\frac{\tau}{2} \,\,\, ,\,\,\, 
\xi' =\frac{\xi}{\sqrt{2}\cos\frac{\tau}{2}}\,\,\, , \,\,\,
\overline{\xi}'=\frac{\overline{\xi}}{\sqrt{2}\cos\frac{\tau}{2}}.
\end{equation}
satisfy (\ref{eq:N=2constraints}). 
The super-Schwarzian transforms according to
the chain rule
\cite{Fu:2016vas} 
\begin{equation}\label{eq:N=2chain_rule}
    \text{Sch}\left[\tau',\xi',\overline{\xi}';
      u,\theta,\overline{\theta}\right]=\left(D_{\theta}\xi\right)
    \left(\overline{D}_{\overline{\theta}}\overline{\xi}\right)\text{Sch}
\left[\tau',\xi',\overline{\xi}';\tau,\xi,\overline{\xi}\right]
+\text{Sch}\left[\tau,\xi,\overline{\xi};u,\theta,\overline{\theta}\right] .
  \end{equation}
This allows
us to express the effective action in terms of fluctuations around the
zero temperature solution $\left( \tau, \xi , \overline{\xi}\right) =
\left( u, \theta, \overline{\theta}\right)$. Fluctuations are described by
two bosonic fields $\epsilon\left( u\right)$ and $\sigma\left( u\right)$
and two fermionic fields $\eta\left( u\right)$ and
$\overline{\eta}\left( u \right)$. Solving (\ref{eq:N=2constraints}) up to
quadratic order in the fluctuations yields \footnote[1]{This expression is
  obtained by considering the infinitesimal limit of 
  reparametrisation given in \cite{Fu:2016vas} in terms of independent
  fields.} 
\begin{equation}\label{eq:N=2rep_zerotemp1}
\tau(u)=u+\epsilon(u)+\theta\overline{\eta}(u)\left[1+\dot{\epsilon}(u)\right]+\overline{\theta}\eta(u)\left[1+\dot{\epsilon}(u)\right]+\theta\overline{\theta}\left[\eta(u)\dot{\overline{\eta}}(u)-\overline{\eta}(u)\dot{\eta}(u)\right],
\end{equation}
\begin{equation}\label{eq:N=2rep_zerotemp2}
\begin{split}
   &\xi(u)=\eta(u)\left[1+\text{i}\sigma(u)+\frac{1}{2}\dot{\epsilon}(u)\right]+\theta\bigg\{1+\overline{\eta}(u)\dot{\eta}(u)+\text{i}\sigma(u)-\frac{1}{2}\sigma^2(u)-\frac{1}{8}\dot{\epsilon}^2(u)+\\
    &+\frac{1}{2}\dot{\epsilon}(u)\left[1+\text{i}\sigma(u)\right]\bigg\}+\theta\overline{\theta}\left\{\dot{\eta}(u)\left[1+\text{i}\sigma(u)+\frac{1}{2}\dot{\epsilon}(u)\right]+\eta(u)\left[\text{i}\dot{\sigma}(u)+\frac{1}{2}\ddot{\epsilon}(u)\right]\right\},
\end{split}   
\end{equation}
\vspace{1mm}
\begin{equation}\label{eq:N=2rep_zerotemp3}
\begin{split}
    \overline{\xi}(u)=\overline{\eta}(u)\left[1-\text{i}\sigma(u)+\frac{1}{2}\dot{\epsilon}(u)\right]+\overline{\theta}\bigg\{1+\eta(u)\dot{\overline{\eta}}(u)-\text{i}\sigma(u)-\frac{1}{2}\sigma^2(u)-\frac{1}{8}\dot{\epsilon}^2(u)+\\
    +\frac{1}{2}\dot{\epsilon}(u)\left[1-\text{i}\sigma(u)\right]\bigg\}-\theta\overline{\theta}\left\{\dot{\overline{\eta}}(u)\left[1-\text{i}\sigma(u)+\frac{1}{2}\dot{\epsilon}(u)\right]+\overline{\eta}(u)\left[-\text{i}\dot{\sigma}(u)+\frac{1}{2}\ddot{\epsilon}(u)\right]\right\}.
\end{split}
\end{equation}
The field $\epsilon(u)$, which also appears in the purely bosonic JT
gravity, is interpreted as a boundary graviton
\cite{Maldacena:2016upp}; the fermionic fields $\eta(u)$ and
$\overline{\eta}(u)$ are boundary gravitinos; finally, $\sigma(u)$
represents the boundary degree of freedom of a graviphoton
\cite{Forste:2020xwx}.  
We can now express the effective action for finite temperature 
super-reparametrisations in terms of this gravity supermultiplet. 
Neglecting total derivatives and
contributions of higher than quadratic order in the fluctuations, we obtain
\begin{align}
    S^{\mathcal{N}=2}_{\text{sch},\,\beta=2\pi}& =\kappa\int du
d\theta d\overline{\theta}\,\,
\text{Sch}\left[\tau',\xi',\overline{\xi}';
u,\theta,\overline{\theta}\right] \nonumber \\
& = C\int_0^{2\pi} du\,\,\left\{
\frac{1}{2}\left[\ddot{\epsilon}^2(u)-\dot{\epsilon}^2(u)\right] 
+2\dot{\sigma}^2(u)
+\left[\eta(u)\dot{\overline{\eta}}(u)
-4\dot{\eta}(u)\ddot{\overline{\eta}}(u)\right]\right\}.      \label{eq:action_fluc}
\end{align}
Here, we have used (\ref{eq:thermal_rep}), (\ref{eq:N=2chain_rule})
and (\ref{eq:N=2rep_zerotemp1}), (\ref{eq:N=2rep_zerotemp2}),
(\ref{eq:N=2rep_zerotemp3}).
This action has been given already in  \cite{Peng:2020euz}.
The form of the $\epsilon$ dependent
contribution coincides
with the one obtained in the purely bosonic setting
\cite{Maldacena:2016upp}. 
\section{Propagators of the fluctuations}\label{sec:propagators}
The propagator for
$\epsilon$ was already obtained in \cite{Maldacena:2016upp}. 
Here, we closely follow the derivation detailed in
\cite{Sarosi:2017ykf}.  In a thermal field theory, time $u$  is
periodic. In our case, the period is $\beta = 2\pi$.
We start by expanding fluctuations into Fourier series,
\begin{align}
  \epsilon(u)=\sum_{n\in\mathbb{Z}}\epsilon_n \text{e}^{\text{i}nu} & ,\,\,\,
  \sigma\left( u\right) = \sum_{n\in{\mathbb Z}}
  \sigma_n\text{e}^{\text{i}nu}  ,\nonumber \\
  \eta(u)=\sum_{m\in\mathbb{Z}+\frac{1}{2}}\eta_m \text{e}^{\text{i}mu}
  &,\,\,\,  \overline{\eta}(u)=\sum_{m\in\mathbb{Z}+\frac{1}{2}}\overline{\eta}_m
  \text{e}^{-\text{i}mu},
\label{eq:fluc_expansion}  
\end{align}
where we have imposed periodic and anti-periodic boundary
conditions on bosonic and fermionic fields, respectively \cite{Dashen:1975xh}.
Plugging \eqref{eq:fluc_expansion} into \eqref{eq:action_fluc} gives 
\begin{equation}\label{eq:action_fluc__n}
  S^{\mathcal{N}=2}_{\text{sch},\,\beta=2\pi}  = \pi C\left\{
    \sum_{n\in\mathbb{Z}}\left[(n^4-n^2)\epsilon_n\epsilon_{-n}+ 2 n^2
    \sigma_n\sigma_n \right] + \sum_{m\in {\mathbb Z} + 1/2} m\left(
    m^2 - \frac{1}{4}\right) \eta_m\overline{\eta}_{m}\right\}.
  \end{equation}
  In Fourier space, the propagators are
 \begin{equation}
    \langle \epsilon_n\epsilon_{k}\rangle=\frac{\delta_{n+k}}{\pi
      C}\frac{1}{n^2(n^2-1)},\,\,\,\,\,\,
    \langle \sigma_n\sigma_k\rangle = \frac{\delta_{n+k}}{4\pi
      C}\frac{1}{n^2}, \,\,\, \,\,\,
    \langle \eta_m \overline{\eta}_k\rangle =\frac{\delta_{m-k}}{2 \pi
      \text{i} C}\frac{1}{m\left( 4 m^2 -1\right)} .
  \end{equation}
Divergencies due to zeros in the denominators are due to zero
modes. These are removed by gauge fixing the supersymmetric extension
of SL$\left( 2, {\mathbb R}\right)$ called SU$\left(\left. 1,1\right|
  1\right)$. This is achieved by just not summing over the
associated Fourier modes when transforming back to position
space. The corresponding sums can be expressed by contour integrals,
\begin{align}
\langle\epsilon(u)\epsilon(0)\rangle &=  \frac{1}{\pi C} \oint_{{\cal C}_\epsilon}
dz\,\,\frac{1}{\text{e}^{2\pi \text{i}z}-1}
\frac{\text{e}^{izu}}{z^2(z^2-1)}, \,\,\, 
\,\,\,
\langle \sigma(u)\sigma(0)\rangle=\frac{1}{4\pi C}
\oint_{{\cal C}_\sigma} dz\,\,
\frac{1}{\text{e}^{2\pi \text{i}z}-1}\frac{\text{e}^{izu}}{z^2}
\nonumber \\
\langle \eta(u)\overline{\eta}(0)\rangle &
=\frac{1}{4\pi \text{i} C}\oint_{{\cal C}_\eta} dz\,\, 
\frac{1}{\text{e}^{4\pi \text{i} z}-1}\frac{\text{e}^{\text{i}zu}}
{z\left(z^2-\frac{1}{4}\right)}, 
\end{align}  
where the integration contours are sets of small circles running
counter clockwise around integer $z$'s excluding zeros in the
denominators in the second factors of the corresponding integrand.
By contour deformation the integrals can be related to integrals in
which the contour runs around the previously excluded poles.  
These can be  evaluated by means of the residue formula. One obtains
\begin{eqnarray}
\lefteqn{\langle\epsilon(u)\epsilon(u')\rangle  =}
\nonumber \\ & & \frac{1}{2\pi C}\!\left[ -\frac{(\lvert
u-u'\rvert -\pi)^2}{2}+(\lvert u-u'\rvert -\pi)\sin\lvert
u - u' \rvert
+1+\frac{\pi^2}{6}+ \frac{5}{2}\cos\lvert u -  u'\rvert \right],  
\label{eq:propagator_epsilon}  
\end{eqnarray}
 \begin{equation} \label{eq:propagator_sigma} 
 \langle \sigma(u)\sigma(u')\rangle  = \frac{1}{48\pi C}\left[3\lvert
  u-u'\rvert^2-6\pi\lvert u-u'\rvert +2\pi^2 \right],  
 \end{equation}
 \begin{equation}\label{eq:propagator_eta} 
 \langle \eta(u)\overline{\eta}(u')\rangle = \frac{1}{4\pi C}\left[(2\pi
  -\lvert u-u'\rvert)\cos\frac{\lvert u-u'\rvert}{2}+3\sin\frac{\lvert
  u-u'\rvert}{2}\right] ,
\end{equation}  
where we have used shift symmetry in $u$ to obtain the dependence on
the second argument $u^\prime$.
\section{Coupling to matter}\label{sec:N=2matter}
Superconformal invariance fixes the form of correlators. For instance,
the two-point function for a superconformal primary $\mathcal{O}$ of
dimension $\Delta$ has the form 
\begin{equation}
    \langle
    \mathcal{O}(u_1)\overline{\mathcal{O}}(u_2)
\rangle\sim\frac{1}{\Xi^{2\Delta}}, 
\end{equation}
where $\Xi$ is invariant under supertranslations. 
Furthermore, the two-point function of chiral (anti-chiral)
superfields is also chiral (anti-chiral).  Thus, if we are interested
in the correlators for chiral and anti-chiral matter fields in
$\mathcal{N}=2$ supersymmetry, $\Xi$ does not only have to be
invariant under the super-translations \eqref{eq:N=2translations} but
also has to be chiral in one coordinate and anti-chiral in the
other. The correct denominator that satisfies these requirements is
\cite{Bulycheva:2018qcp} 
\begin{equation}\label{eq:N=2inv_dist}
    \Xi=u_1-u_2-2\overline{\theta}_1\theta_2-\theta_1\overline{\theta}_1-\theta_2\overline{\theta}_2.
\end{equation}
Following the general rules of AdS/CFT duality, superconformal symmetry
fixes the on-shell value of the bulk matter action \cite{Witten:1998qj}
\begin{equation}\label{eq:startingN=2_2pointfunc}
    S_{\text{matter}}=
\mu\int d\tau'_1 d\tau'_2 d\xi'_2 d\overline{\xi'}_1\,\, 
\frac{\Psi(\tau'_2,\xi'_2)\overline{\Psi}(\tau'_1,\overline{\xi'}_1)}
{\left|\tau'_1-\tau'_2-2\overline{\xi'}_1\xi'_2-\xi'_1\overline{\xi'}_1
-\xi'_2\overline{\xi'}_2\right|^{2\Delta}},
\end{equation}
where $\mu$ is some constant and $\Psi(\tau'_2,\xi'_2)$,
$\overline{\Psi}(\tau'_1,\overline{\xi'}_1)$ are respectively chiral and
anti-chiral matter multiplets corresponding to the boundary values of fields
living in the bulk of nAdS$_2$. On the CFT side, these are sources
coupling to superconformal primaries of dimension $\Delta$. 
The zero temperature correlator is obtained by taking variational
derivatives of the exponential of $S_{\text{matter}}$ with respect to
  $\Psi\left( \tau'_i , \xi'_i, \overline{\xi}'_i\right)$ and 
$\overline{\Psi}\left(
    \tau'_j , \xi'_j ,\overline{\xi}'_j\right)$ and viewing $\tau', \xi' ,
  \overline{\xi}'$ as superspace coordinates. The finite temperature
  correlator follows from variational derivatives with respect to
  superfields depending on the unprimed coordinates
  $\tau,\xi,\overline{\xi}$ where primed and unprimed quantities are
  related as in (\ref{eq:thermal_rep}). Finally, corrections due to
  broken superconformal invariance can be taken into account when
  considering variational derivatives with respect to superfields
  depending on $u,\theta,\overline{\theta}$ and the relation to $\tau, \xi,
  \overline{\xi}$ is given in (\ref{eq:N=2rep_zerotemp1}),
  (\ref{eq:N=2rep_zerotemp2}), (\ref{eq:N=2rep_zerotemp3}). In order
  to take the variational derivatives with respect to fields depending
  on $u,\theta,\overline{\theta}$ we use the fact that they are related by
  super-reparametrisations to $\tau',\theta',\overline{\theta}'$. There will
  be a Berezian from the superspace measure \cite{Fu:2016vas} whereas
  the transformation properties of the superfields $\Psi$ and
  $\overline{\Psi}$ follow from the AdS/CFT dictionary equating them with
  currents coupling to superconformal primaries. Taking all these
  contributions into account we obtain
  \begin{equation}\label{eq:N=2_matter}
  S^{\mathcal{N}=2}_{\text{matter}} = \mu \int d{u_{ac}}_1d\overline{\theta}_1
  d{u_{c}}_2d\theta_2  \,\,
  \frac{\Big(
          D_{\overline{\theta}_2}\overline{\xi}'_2\Big)^{2\Delta}\Big( 
      D_{\theta_1}\xi'_1\Big)^{2\Delta} \Psi\left({u_{c}}_2,\theta_2\right)
  \overline{\Psi}\left({u_{ac}}_1,\overline{\theta}_1\right)}
{\left|\tau'_1-\tau'_2-2\overline{\xi'}_1\xi'_2-\xi'_1\overline{\xi'}_1
-\xi'_2\overline{\xi'}_2\right|^{2\Delta}},   
\end{equation}
where the primed quantities are viewed as functions of
$u,\theta,\overline{\theta}$  with the corresponding label.
Further,  $u_{c} = u  - \theta\overline{\theta}$ and $u_{ac} = u +
\theta\overline{\theta}$ are the chiral and anti-chiral combinations on
which associated superfields depend.
\section{Four-point functions}\label{sec:N=2_4pointfunc}
In this section we compute the four-point functions at finite
temperature for operators  $V$ and
$\overline{V}$ of conformal dimension $\Delta$, dual to the
matter fields $\overline{\Psi}$ and $\Psi$. Four-point functions at
zero temperature were already computed in \cite{Forste:2020xwx}.
It is useful to introduce a second pair of superconformal primaries
$W$ and $\overline{W}$ which have vanishing two-point functions with
$\overline{V}$ and $V$. On the supergravity side this corresponds to adding
$\tilde{S}_{\text{matter}}$ which is equal to $S_{\text{matter}}$ but where
the  matter multiplets are labelled with an additional tilde. 
The generating functional of  correlators for superconformal operators of
the boundary theory coincides with the partition function of the gravity
theory in which the currents are identified with boundary values of
matter superfields
\begin{equation}
\begin{gathered}
Z\left[ \Psi,\overline{\Psi},\tilde{\Psi},\tilde{\overline{\Psi}}\right] =
  \Big\langle \exp\left[\int du d\theta 
  d\overline{\theta}\left(\Psi\overline{V}+\overline{\Psi}V + \tilde{\Psi}\overline{W} +
    \tilde{\overline{\Psi}} W \right)\right]
\Big\rangle_{\text{SCFT}}\\\stackrel{\text{AdS/CFT}}{=} 
\int{\cal D}^\prime\epsilon {\cal D}^\prime \sigma
{\cal D}^\prime \eta{\cal D}^\prime \overline{\eta} \,\,
\text{e}^{-S^{\mathcal{N}=2}_{\text{sch},\,
    \beta=2\pi}-S^{\mathcal{N}=2}_{\text{matter}}
  -\tilde{S}^{\mathcal{N}=2}_{\text{matter}
  }} .
\end{gathered}
\end{equation}
From the generating functional one can e.g.\ obtain four-point
functions by taking variational derivaties,
\begin{equation}\label{eq:starting_4pt}
    \langle{\cal T} \overline{V}V\overline{W}W\rangle=\frac{\delta^4
      Z}{\delta
      \Psi\delta\overline{\Psi}\delta\tilde{\Psi}\delta\tilde{\overline{\Psi}}}
    \bigg\rvert_{\Psi,\overline{\Psi},\tilde{\Psi},\tilde{\overline{\Psi}}=0} =
\left(\frac{\delta^2
      Z}{\delta
      \Psi\delta\overline{\Psi}}\right)
  \left(\frac{\delta^2
      Z}{
      \delta\tilde{\Psi}\delta\tilde{\overline{\Psi}}}\right)
  \bigg\rvert_{\Psi,\overline{\Psi}
    \tilde{\Psi},\tilde{\overline{\Psi}}=0}. 
\end{equation}
where ${\cal T}$ denotes time ordering. 
\newpage 
\noindent One finds
\begin{multline}
\langle {\cal T} \overline{V}\left( u_{ac1} ,\overline{\theta}_1\right)V\left(
  u_{c2},\theta_2 \right) \overline{W}\left( u_{ac3},\overline{\theta}_3\right)
W\left( u_{c4}, \theta_4\right)\rangle   \\
=\int{\cal D}^\prime\epsilon {\cal D}^\prime \sigma
{\cal D}^\prime \eta{\cal D}^\prime \overline{\eta}\,\,
N J\left({u_{ac}}_1,{{u_c}}_2\right)J\left({u_{ac}}_3,{{u_c}}_4\right) 
\text{e}^{-S_{\text{Schw}}}, 
\label{eq:rhs_4pt}
\end{multline}
where
\begin{equation}
  N^{-1} = 4^{2\Delta}\sin^{2\Delta}
  \frac{{u_{ac}}_1-{{u_c}}_2}{2}\sin^{2\Delta}\frac{{u_{ac}}_3-{{u_c}}_4}{2}
\end{equation}
and
\begin{equation}
  J\!\left({u_{ac}}_1,\overline{\theta}_1,{{u_c}}_2,
    \theta_2\right)=A\!\left({u_{ac}}_1,{{u_c}}_2\right) 
  +B\!\left({u_{ac}}_1,{{u_c}}_2\right)\overline{\theta}_1
  +C\!\left({u_{ac}}_1,{{u_c}}_2\right)\theta_2
  +D\!\left({u_{ac}}_1,{{u_c}}_2\right)\overline{\theta}_1\theta_2  
\end{equation}
and the coefficients can be read off \eqref{eq:N=2_matter}.
Dropping the $a$ and $ac$ labels and introducing the convenient
notation $u_{ij}=u_i-u_j$, we find
\begin{align}
   A\left(u_1,u_2\right) 
= & 1+\Delta\left[\frac{-\epsilon\left(u_1\right)
+\epsilon\left(u_2\right)}{\tan\frac{u_{12}}{2}}+\dot{\epsilon}(u_1)
+\dot{\epsilon}(u_2)+2\text{i}\left(\sigma(u_1)-\sigma(u_2)\right)\right], \\
B\left(u_1,u_2\right) = & 2\Delta\left[\frac{\eta(u_1)}{\tan\frac{u_{12}}{2}}
-\frac{\eta(u_2)}{\sin\frac{u_{12}}{2}}-2\dot{\eta}(u_1)\right], \\
C\left(u_1,u_2\right) =&2\Delta\left[\frac{\overline{\eta}(u_1)}
{\sin\frac{u_{12}}{2}}-\frac{\overline{\eta}(u_2)}
{\tan\frac{u_{12}}{2}}-2\dot{\overline{\eta}}(u_2)\right],\\
D\left(u_1,u_2\right)=&\frac{2\Delta}{\sin\frac{u_{12}}{2}}
\Big[1-(1+2\Delta)\frac{\epsilon(u_1)-\epsilon(u_2)}
{2\tan\frac{u_{12}}{2}}+(1+2\Delta)
\left(\dot{\epsilon}(u_1)+\dot{\epsilon}(u_2)\right) \nonumber\\
  & +2\text{i}
\left(-1+2\Delta\right)\left(\sigma(u_1)-\sigma(u_2)\right)\Big].
\end{align}
We decompose the superconformal primaries as
\begin{equation}
\begin{gathered}
  V\left(u_2,\theta_2\right)=V_{\theta}\left({{u_c}}_2\right)
  \theta_2+V_{\phi}\left({{u_c}}_2\right)\equiv 
    V_{\theta_2}\theta_2+V_{\phi_2}\\ 
    \overline{V}\left(u_1,\theta_1\right)=\overline{V}_{\overline{\theta}}\left(
        {u_{ac}}_1\right)\overline{\theta}_1+\overline{V}_{\overline{\phi}}\left(
        {u_{ac}}_1\right)\equiv\overline{V}_{\overline{\theta}_1}
      \overline{\theta}_1+\overline{V}_{\overline{\phi}_1},
\end{gathered}
\end{equation}
with similar expressions holding for $W$, $\overline{W}$. 
By comparing both sides of \eqref{eq:starting_4pt} we are then led to
the following non-vanishing connected four-point functions,
\begin{align}
   \langle {\cal T}\overline{V}_{\overline{\phi}_1}
  V_{\phi_2}\overline{W}_{\overline{\phi}_3}W_{\phi_4}\rangle=& N \wick{\c A(u_1,u_2) \c
 A(u_3,u_4)},  \label{eq:ppbz}\\
\langle{\cal T}
  \overline{V}_{\overline{\theta_1}}V_{\theta_2}\overline{W}_{\overline{\phi}_3}W_{\phi_4}
\rangle=&
N \wick{\c D(u_1,u_2) \c A(u_3,u_4)},
\label{eq:bzpp}\\
\langle{\cal T}\overline{V}_{\overline{\theta}_1}V_{\theta_2}
\overline{W}_{\overline{\theta}_3}
W_{\theta_4}\rangle= & 
N \wick{\c D(u_1,u_2) \c D(u_3,u_4)},                           
\label{eq:bzbz}\\
\langle{\cal T}
  \overline{V}_{\overline{\theta}_1}V_{\phi_2}\overline{W}_{\overline{\phi}_3}
W_{\theta_4}\rangle
  = & -N\wick{\c B(u_1,u_2) \c C(u_3,u_4)}, 
\label{eq:bppz}\\
\langle{\cal T}\overline{V}_{\overline{\phi}_1}V_{\theta_2}\overline{W}_{\overline{\theta}_3}
W_{\phi_4}\rangle = & -N\wick{\c C(u_1,u_2) 
\c B(u_3,u_4)}.                                              
\label{eq:pzbp}
\end{align}
Horizontal brackets denote Wick contractions of super-Schwarzian
fluctuations: only contributions quadratic in fluctuations are
contained in the  connected four-point function.  The pair of
fluctuations is replaced by its 
propagator. Since the propagators depend on absolute values of their
arguments the results depend on the chosen ordering. The arrangement
which, upon canonical continuation to Minkowski time, leads to out
of time order correlators (OTOCs) is $u_1> u_3 > u_2 > u_4$. Before
continuation to Minkowski time one obtains
\begin{align}
 \langle \overline{V}_{\overline{\phi}_1}\overline{W}_{\overline{\phi}_3}
  V_{\phi_2}W_{\phi_4}\rangle=&  \frac{N\Delta^2}{2\pi C}\bigg[
\left(\frac{u_{12}}{\tan\frac{u_{12}}{2}}-2\right)
\left(\frac{u_{34}}{\tan\frac{u_{34}}{2}}-2\right)
     +2\pi \frac{\sin \frac{u_{12}+u_{34}}{2}
-\sin\frac{u_{23}+u_{14}}{2}}{\sin\frac{u_{12}}{2}
\sin\frac{u_{34}}{2}}\nonumber\\ &
+2\pi\frac{u_{23}}{\tan\frac{u_{12}}{2}\tan\frac{u_{34}}{2}}
+u_{12}u_{34}+2\pi u_{23}\bigg],\label{eq:eOTOC1}\\ 
\langle\overline{V}_{\overline{\theta_1}}\overline{W}_{\overline{\phi}_3}
V_{\theta_2}W_{\phi_4}\rangle  =&-\frac{N\Delta^2}{2\pi C}\frac{1}{\sin\frac{u_{34}}{2}}\bigg \{ (2\Delta+1)\bigg[\left(\frac{u_{12}}{\tan\frac{u_{12}}{2}}-2\right)
\left(\frac{u_{34}}{\tan\frac{u_{34}}{2}}-2\right)\nonumber\\
 &+2\pi \frac{\sin \frac{u_{12}+u_{34}}{2}
-\sin\frac{u_{23}+u_{14}}{2}}{\sin\frac{u_{12}}{2}
\sin\frac{u_{34}}{2}}
\bigg]+(2\Delta-1)\left[u_{12}u_{34}+2\pi u_{23}\right]\bigg\},\\
\langle\overline{V}_{\overline{\theta}_1} 
\overline{W}_{\overline{\theta}_3} V_{\theta_2} 
W_{\theta_4}\rangle =&\frac{N\Delta^2}{2\pi C}\frac{1}{\sin\frac{u_{12}}{2}\sin\frac{u_{34}}{2}}\bigg \{ (2\Delta+1)^2\bigg[\left(\frac{u_{12}}{\tan\frac{u_{12}}{2}}-2\right)
\left(\frac{u_{34}}{\tan\frac{u_{34}}{2}}-2\right)\nonumber\\
 &+2\pi \frac{\sin \frac{u_{12}+u_{34}}{2}
-\sin\frac{u_{23}+u_{14}}{2}}{\sin\frac{u_{12}}{2}
\sin\frac{u_{34}}{2}}
\bigg]+(2\Delta-1)^2\left[u_{12}u_{34}+2\pi u_{23}\right]\bigg\},\\
\langle
  \overline{V}_{\overline{\theta}_1}\overline{W}_{\overline{\phi}_3}
V_{\phi_2}
W_{\theta_4}\rangle
  = &  -\frac{2N\Delta^2}{\pi C}\frac{1}{\sin
      \frac{u_{12}}{2}\sin\frac{u_{34}}{2}}\left[u_{23}
\cos\frac{u_{23}}{2}-3\sin\frac{u_{23}}{2}\right], \\
\langle\overline{V}_{\overline{\phi}_1}\overline{W}_{\overline{\theta}_3}
V_{\theta_2}
W_{\phi_4}\rangle = &-\frac{2N\Delta^2}{\pi C}\frac{1}{\sin 
\frac{u_{12}}{2}\sin\frac{u_{34}}{2}}\Big[u_{32}
\cos\frac{u_{14}}{2}+\sin\frac{u_{14}}{2}
\nonumber\\ & \quad
              -\sin\frac{u_{12}+u_{42}}{2}+\sin\frac{u_{13}+u_{43}}{2}\Big] .
\label{eq:eOTOC4}
\end{align}
To obtain the real time OTOCs we proceed by analytic continuation 
\cite{Maldacena:2016upp}. Reinstalling also temperature dependence we
replace
\begin{equation}
u_1 = \frac{2\pi\text{i}}{\beta} a\,\,\, ,\,\,\, u_2 = 0\,\,\, ,\,\,\,
u_3=\frac{2\pi\text{i}}{\beta}\left( b +\hat{u}\right) \,\,\, ,\,\,\, 
u_4 =  \frac{2\pi\text{i}}{\beta} \hat{u} 
\end{equation}
in equations (\ref{eq:eOTOC1})-(\ref{eq:eOTOC4}). 
Furthermore, one is interested in the region with $a,b$ of the order of
the dissipation time $\beta$ whereas $\hat{u}$ is much larger than the
dissipation time while still much shorter than the scrambling time, i.e.\
$\frac{2 \pi }{\beta} \ll \hat{u} \ll \frac{2 \pi }{\beta}\log
\frac{C}{\beta}$. Keeping only the leading contributions (growing
exponentially with $\hat{u}$) one finds 
\begin{align}
  \langle
 \overline{V}_{\overline{\phi}}(0)\overline{W}_{\overline{\phi}}\left(
  \hat{u}\right) V_{\phi}(0)
  W_{\phi}(\hat{u}) 
  \rangle_{\beta} & \sim
  \frac{\Delta^2}{C}\text{e}^{\frac{2\pi}{\beta}\hat{u}},\\
   \langle
  \overline{V}_{\overline{\theta}}(0)\overline{W}_{\overline{\theta}}\left(
  \hat{u}\right)V_{\theta}(0)
  W_{\theta}(\hat{u})  
  \rangle_{\beta} & \sim
  \frac{\Delta^2}{C}\left(1+2\Delta\right)^2
\text{e}^{\frac{2\pi}{\beta}\hat{u}},\\
 \langle
  \overline{V}_{\overline{\phi}}(0)
\overline{W}_{\overline{\theta}}\left( \hat{u} \right) V_{\phi}(0)
  W_{\theta}(\hat{u}) 
  \rangle_{\beta} & \sim
  \frac{\Delta^2}{C}\left(1+2\Delta\right)\text{e}^{\frac{2\pi}{\beta}\hat{u}},
\\ 
 \langle
  \overline{V}_{\overline{\theta}}(0)\overline{W}_{\overline{\phi}}\left(
  \hat{u})\right)
V_{\theta}(0)
  W_{\phi}(\hat{u})
  \rangle_{\beta}& \sim
   \frac{\Delta^2}{C}\left(1+2\Delta\right)
                   \text{e}^{\frac{2\pi}{\beta}\hat{u}}.
\end{align}
The exponential growth of the above correlators originates from
graviton ($\epsilon$) exchange. The corresponding Lyapunov exponent is
\begin{equation}
  \lambda^{(2)} = \frac{2\pi}{\beta},
\end{equation}
i.e.\ it saturates the bound of \cite{Maldacena:2015waa}. The
superscript two indicates that this behaviour is due to graviton (spin
two particle) exchange. 
Graviton exchange does not contribute to the four-point functions
(\ref{eq:bppz}) and (\ref{eq:pzbp}) at the given order in
  $\frac{1}{C}$. For those one finds from gravitini exchange 
\begin{equation}
    \langle
    \overline{V}_{\overline{\phi}}(0)\overline{W}_{\overline{\theta}}(\hat{u})
V_{\theta}(0) 
    W_{\phi}(\hat{u}) 
    \rangle_{\beta}\sim\langle
    \overline{V}_{\overline{\theta}}(0)
\overline{W}_{\overline{\phi}}(\hat{u})V_{\phi}(0) 
    W_{\theta}(\hat{u})
    \rangle_{\beta}\sim
    \frac{\Delta}{C}\hat{u}\text{e}^{\frac{\pi}{\beta}\hat{u}}, 
\label{eq:gravitino}
\end{equation}  
implying a Lyapunov exponent
\begin{equation}
  \lambda^{(3/2)} = \frac{\pi}{\beta}.
\label{eq:la-spin}
\end{equation}
This agrees with the pattern found in the context of ${\cal N}=1$
supersymmetry \cite{Perlmutter:2016pkf ,Narayan:2017hvh},
\begin{equation}
  \lambda^{(s)} = \frac{2\pi}{\beta}\left( s - 1\right) ,
\label{eq:la-spinLit}
\end{equation}
where $s$ is the spin of the gravitational mode being exchanged. From
(\ref{eq:propagator_sigma})  we observe that
combinations to which only the exchange of a graviphoton $\sigma$
(spin one)
contributes do not show exponential behaviour in agreement with
(\ref{eq:la-spinLit}). Notice also that for the result (\ref{eq:gravitino})
the introduction of the second primary $W$ with vanishing two-point
function, $\langle W\overline{V}\rangle =0$, is crucial. If e.g.\ we had
considered the 4-point functions with $V$, $\overline{V}$'s only,
there would be additional terms with
$$J\left( u_{ac1}, u_{c4}\right) J\left( u_{ac3}, u_{c2}\right)  $$
in (\ref{eq:rhs_4pt}).
These would result in graviton exchange
contributing to 
(\ref{eq:gravitino}) with $W$, $\overline{W}$ replaced by $V$,
$\overline{V}$. Then also these components would show exponential
growth with maximal Lyapunov exponent.
\section{Summary and Outlook}

In this paper we studied the $\mathcal{N}=2$ supersymmetric
Jackiw-Teitelboim theory at finite temperature and computed the
Lyapunov exponents. We found that their value depends on the
particular component of the four-point function within  the
supermultiplet.
There are compenents with maximal Lyapunov exponent $2\pi/\beta$. For
them the leading correction contains graviton exchange. For other
components the graviton does not couple to corresponding
bi-locals. Due to gravitini exchange the leading contribution yields a
Lyapunov exponent $\pi/\beta$. One could also consider superpositions
with graviphoton exchange only. They do not show exponential time
dependence (chaotic behaviour). Finally, we argued that for non
vanishing two-point function, $\langle \overline{V}W\rangle \not= 0$,
all components have maximal Lyapunov exponent.

The ${\cal N}=2$ case is interesting. For instance, it is related to four
dimensionial 1/4 BPS black holes
\cite{Cacciatori:2009iz,Hristov:2010ri,Forste:2020xwx}, whereas, to
our knowledge, a
higher dimensional solution leading to ${\cal N} =1$ JT gravity in its
near horizon limit has not been identified so far.   

Several generalisations of the  present calculation are possible. One could add
flavour symmetry as in \cite{Narayan:2017hvh}, or consider higher
amounts of sypersymmetry, e.g. $N=3,4$ super-Schwarzian actions. Furthermore, subleading corrections in $1/C$ 
might be interesting. For the bosonic case these have been analysed in  
\cite{Qi:2019gny}, for ${\cal N}=1$ some contributions are given in
\cite{Mertens:2020pfe}.

\section*{Acknowledgements}  
This work was supported by ``Bonn-Cologne
Graduate School for Physics and Astronomy'' (BCGS).

\bibliographystyle{JHEP}
\bibliography{references}

\end{document}